# A Controller for Network-Assisted CACC based Platooning

Sanket Partani
*University of Kaiserslautern*
Kaiserslautern, Germany
partani@eit.uni-kl.de

Andreas Weinand
*University of Kaiserslautern*
Kaiserslautern, Germany
weinand@eit.uni-kl.de

Hans D. Schotten
*University of Kaiserslautern*
*German Research Center*
*for Artificial Intelligence*
Kaiserslautern, Germany
schotten@eit.uni-kl.de

*Abstract*— Platooning involves a set of vehicles moving in a cooperative fashion at equal inter-vehicular distances. Taking advantage of wireless communication technology, this paper aims to show the impact of network protocols on a platoon using a controller, based on the Cooperative Adaptive Cruise Control (CACC) principles. The network protocols used in this work are DSRC (Dedicated Short Range Communication) and LTE-V2V sidelink (Mode 4). The main focus of this work is to showcase the ability of the controller to maintain platoon stability despite having uncertainties in both, the platoon and the message delivery rates over the network protocols. The controller interacts with all vehicles using messages transmitted over the network protocols. The controller is designed to be responsible for micro-managing every vehicle in the platoon and to ensure that the platoon does not break under any circumstances. SUMO (Simulation of Urban MObility) is used as the simulation platform. Results indicate, that the controller manages to achieve platoon stability in all scenarios, unless a set number of consecutive messages are not transmitted, in which case it leads to collisions. This work also presents certain bottlenecks pertaining to wireless communication with vehicles.

## I. Introduction

There has been a rapid increase in the number of vehicles in both the private ownership and in the transport sector. This has resulted in overcrowded roads in the cities and on the highways, leading to higher accident rates, fuel consumption and emissions. Hence, traffic management strategies have gained more importance of late. Platooning is one such cooperative strategy, wherein semi-autonomous vehicles drive in almost perfect longitudinal unison. This not only ensures lower fuel consumption, but also increases vehicle throughput on the highways. As much as 6% of fuel savings for the leading vehicle and up to 10% of fuel savings for the following vehicles have been determined in track experiments [1].

Platooning can be traced back to technologies such as Adaptive Cruise Control (ACC), wherein a vehicle decides a time headway (gap) to its preceding vehicle, based on its current speed. This gap can be maintained using LIDAR and radar technologies. Successful examples of platooning implementations are the SARTRE Project [2], Japan's Energy ITS Project [3] and the California PATH Project [4]. With advancement in vehicle-to-vehicle (V2V) communication technologies supporting low latency and high reliability communication, platooning can also be achieved using messages to interact between vehicles. These messages contain information such as vehicle's speed, acceleration, current position, etc. and can help control the vehicles, based on their distances to their preceding vehicles. However, the response of vehicles in the platoon depends on the interval, latency, reliability, etc. of the messages exchanged between them.

Until recently, the V2V communication domain was dominated by the DSRC protocol. Based on the IEEE 802.11 standard, it is used in the 5.9 GHz band in the USA, Europe and Japan. However, recent advancements made in the 3GPP (Release 13) LTE network specifications enable vehicles to interact (technologically and economically viable) with not only other vehicles (V2V), but also with the infrastructure (V2I), pedestrians (V2P), etc. This exchange of information makes the vehicle more reactive to its surroundings and hence, increases safety. Release 15 of 3GPP introduced 5G and various safety use cases along with it, and states various requirements needed to achieve it [5] [8]. It paves the path for the evolution of cellular-vehicle-to-everything (C-V2X) technology [6] [7] [9] with specific use cases such as platooning.

Previous studies on the CACC platooning [10], [11] involved scenarios that showed increase in traffic throughput and savings in fuel consumption with decreasing inter-vehicular distances. These works also showed that CACC can be viably used in the platooning scenario. This work presents a single platoon on a highway scenario. The platoon undergoes various stages – forming, braking and acceleration; thus, providing for enough opportunities for inducing an instability in the platoon. The controller, designed and implemented for the scenario, is responsible for platoon formation and stability. The controller relies solely on messages to interact with its members over the V2V communication link. The controller also ensures that any instability in the platoon is dealt with locally and is not propagated up or down the platoon chain. Messages are exchanged over the DSRC and the LTE-V2V sidelink protocol and the results are compared. SUMO is used to simulate the scenario. SUMO is a microscopic traffic simulator, which with the help of car-following models can help achieve realism in simulations.

The remainder of this work is divided into the following sections. Section 2 discusses platooning and its various control strategies. The controller design and logic is explained in Section 3. Section 4 deals with the V2V network protocols. Section 5 deals with the simulation scenarios. Section 6 discusses the results and compares the controller performance over the two communication protocols. Finally, conclusions and avenues for future research are given in Sections 7 and 8 respectively.

## II. PLATOONING

On highways, where the average speeds of vehicles are higher than 90 kmph, vehicles need more power to overcome aerodynamic drag. At this speed, the drag accounts for 53% of the fuel consumption [12]. Drag depends on the effective frontal area of the vehicle, the speed of the vehicle and the shape of the vehicle. One way to reduce drag is to streamline the vehicle or make the vehicle more aerodynamic. Another way is to reduce distance between two vehicles driving in the same lane. The distance reduction has a two-fold effect. It reduces the wake profile (generated at the back of the vehicle) for the leading vehicle and reduces the front-end drag for the following vehicle. The lower the distance, the higher is this effect of pressure reduction/increase. Adding more vehicles to this system results in a convoy of vehicles, i.e., a platoon.

Once formed, the platoon needs to be controlled. This is done so that vehicles do not remain in a state of disarray, i.e., increase or decrease inter-vehicular distance (IVD) to maintain the platoon. Control can be achieved by either controlling the vehicles' speed (ACC) or the inter-vehicular distance (CACC). With regards to platooning using network-assisted CACC, one of the following controller setups introduced can be used.

### A. Centralized Strategy

As it can be seen in Fig 1, the onus of platoon controller falls on the platoon leader. Each vehicle transmits its current speed, position and acceleration to the leader. And depending on these values and/or the current traffic situation, the leader sends back messages to the apt platoon member to slow down or accelerate, so that the IVD is maintained.

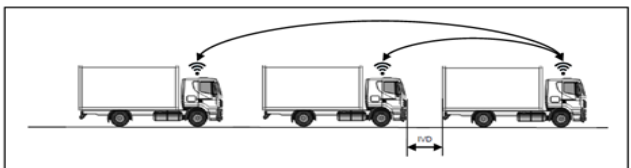

Fig 1: Centralized strategy

However, this strategy has its shortcomings. Depending on the wireless technology used, the platoon length will have to be limited, since there can be messages which would not reach the last member of the platoon.

### B. Decentralized Strategy

In this control strategy, every platoon member takes some initiative to maintain the IVD and hence, the stability of the platoon. This can be envisaged using one of the following ways:

a) vehicle information of only preceding vehicle - strategy revolves around on-board sensor measurements being communicated to the following vehicle

b) vehicle information of preceding and leading vehicle - strategy can be used to achieve longitudinal control (by tracking preceding vehicle) and lateral control (by tracking trajectory of the platoon leader)

The advantage of this system is that there is no limit to the number of vehicles in the platoon. However, the technological and the computational overhead for each vehicle is higher compared to the centralized approach.

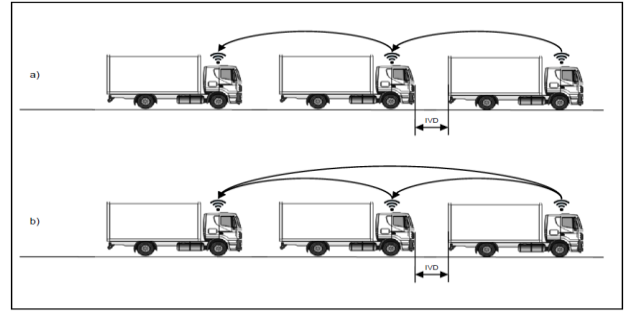

Fig 2: Decentralized strategy

## III. PLATOON CONTROLLER

The controller is designed to handle and test the capabilities of network-assisted platooning. Hence, the controller should be able to form platoons, adapt to the surrounding traffic, and at the same time keep all platoon members stable. Certain assumptions are made whilst developing the controller. Firstly, the platoon moves only in a straight line, i.e., lateral movement of vehicles in the lane is not considered. Secondly, once a vehicle joins the platoon, it never leaves it. And lastly, all vehicular components (CAN bus, Antenna, etc.) work ideally.

### A. Controller Architecture

The top level system architecture is shown in Figure 3. The controller is coded in python to use the TraCI package provided with python.

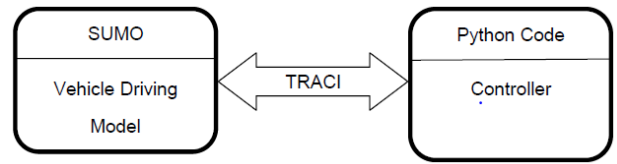

Fig 3: Top-level system architecture

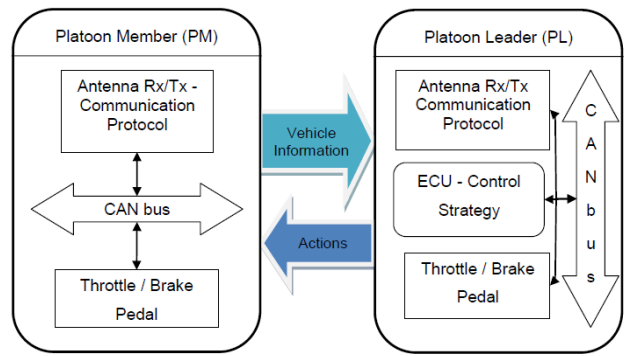

Fig 4: Vehicle architecture

The platoon members and the leader are designed and implemented as shown in Figure 4. The figure also shows the interaction between the leader and the members. A single antenna acts as both, the transmitter and the receiver.

### B. Controller Logic

The controller is setup in the PL (platoon leader) as part of the centralized control strategy. It was chosen over the decentralized approach due to its lower computational complexity.

#### 1) Selecting viable PL

The PL is always chosen to be a truck. This is due to the high frontal area of a truck.

#### 2) Selecting viable PMs

The vehicles are chosen to be PMs (platoon member) if they satisfy the following criteria (in decreasing order of priority):

- Distance from the last PM

Lower distances are preferred.

- Speed of the vehicle

Vehicle is chosen if its current speed is +/- 25% of the current platoon speed.

#### 3) Joining manoevre

The selected vehicle can connect with the platoon at the end, or it can merge into the platoon at a position pre-decided by the controller. This position is decided based on the vehicle's current position.

#### 4) Information Exchange

Information exchange is a key element of the controller. The controller receives the current position (x, y coordinates), current speed and the current IVD to its preceding vehicle. Based on the information provided by all the PMs, the controller sends each individual PM an action message. This message could inform the PM to accelerate, decelerate or maintain its current speed in order to maintain platoon stability.

#### 5) Platoon Stability

The controller is always (unless the message is not received) aware of the IVD of any PM to its preceding vehicle. If the IVD is higher than the desired IVD, the controller commands the concerned PM to accelerate its current velocity by a margin of 0.4 ms$^{-1}$, and vice versa. As this method of vehicle control is equivalent to a saw-tooth controller, there are expected to be high frequency of overshoots and undershoots for the PMs whilst achieving stability. Hence, a tolerance range of +/- 10% of the desired IVD rather than the IVD itself is provided to the controller to limit them.

## IV. V2V PROTOCOLS

We consider high level performance metrics, such as the PDR (packet delivery ratio), in order to model the network protocols. The following performance metrics for DSRC and LTE-V2X based networks were conducted under similar conditions compared to our scenario. This high-level abstraction of the network protocols helps to focus this work on the controller's capabilities to stabilize the platoon. Also, latency, network congestion and interference is not considered.

### A. DSRC

DSRC is primarily designed to support low latency communications for vehicular networking. It is a short to medium range communications technology. It has a similar signal structure to that of Wi-Fi, and to avoid interference with other Wi-Fi devices, it operates in the 5.9 GHz band. The used 75 MHz band comprises of 7 channels (1 control and 6 service) and 1 reserve guard band of 5 MHz. More information on the DSRC protocol can be found in [13].

For the DSRC implementation, the PDR values are taken from [14]. In this experimental setup of 3 GM (General Motor) vehicles with omni-directional roof mounted antennas, a number of VSC (vehicle safety communication) applications are used to determine the reliability of the DSRC protocol.

### B. LTE-V2V sidelink

The LTE-V2V sidelink or Mode 4 was first introduced in the first version of Release 14 by 3GPP in September, 2016. Vehicles create an ad-hoc network to communicate with each other. In Mode-4, vehicles select resources autonomously, which is an upgrade on the DSRC, where the vehicles compete for channel access. Resources are allocated using a semi-persistent scheduling scheme. To avoid packet collisions, vehicles include their packet transmission interval and the value of its reselection counter in the sidelink control information (SCI). More information of Mode 4 can be found in [15].

Similar to the DSRC implementation, for the implementation of LTE-V2V sidelink, we used the PDR statistics from [16]. In this setup, the conditions for PDR to distance calculation are calculated by populating a highway with a set number of vehicles.

## V. SIMULATION SET-UP

### A. SUMO Road Network

The road network for simulations is created using NETEDIT. A single lane highway section of 4 km length is chosen for the scenario. The road speed is limited to 130 kmph.

### B. Vehicle Model

A total of 5 trucks are populated on the road section. These vehicles form the platoon, as directed by the controller. The parameters for trucks are defined in Table 1.

The antennas on all vehicles are simulated to be 0.5 m from the front-end of each vehicle. Also, the vehicle reaction time, i.e., the time from when it receives the action message to the time the pedals are pressed, is simulated to be 300 ms. If a higher vehicle reaction time is chosen, higher frequency in overshooting and undershooting is observed and may result in collisions.

TABLE I. VEHICLE PARAMETERS

| Parameters | Value |
|---|---|
| Length | 13.6 m |
| Max acceleration | 2.5 ms-2 |
| Max deceleration | 10 ms-2 |
| Max speed | 27.77 ms-1 |

*C. SUMO parameters*

To ensure that only the controller has complete command over the platoon and not the car-following model of SUMO, parameters such as the *minGap* and the *reaction time (τ)* for each vehicle are set to 0. Only these parameters decide if the sole control of the vehicles is with the controller or not. When these parameters are chosen differently, SUMO tends to interfere with the controller, which is not desired.

*D. Simulation Scenario*

The overall simulation looks as follows:

- Platoon formation
- Emergency braking
- Acceleration

Once all vehicles join the platoon, the PL is forced to come to an emergency halt. This creates some instability in the platoon. As during breaking, if a PM doesn't receive an action to brake, it will definitely collide into its preceding vehicle. After a short amount of time, the PL accelerates to 80 kmph, and the PMs follow. The vehicles are said to be stable when they are in the desired IVD range. The time taken for the controller to achieve stability, after accelerating back from the stand-still position, for all PMs is the chosen performance indicator for this work. The platoon is tested with an IVD of 5m and a message rate of 80 Hz, 40 Hz and 10 Hz.

## VI. RESULTS

A time step in the simulation domain is assumed to be equal to the message rate (for example, for a message rate of 80Hz, the time step is 12.5 ms. So time step 10000 is equal to 125 seconds). A complete simulation scenario can be seen in Figure 6. All the other figures show only the simulation from the start of the Acceleration phase (Section 5.4), wherein the vehicles resume their route from a point of complete standstill.

When it comes to the scenario of emergency braking, collisions were detected when messages were not transmitted. However, these results are not a reflection of those collision simulations.

From Figures 5 and 6, it can be clearly seen that all PMs are in a constant state of disarray. Localised overshooting and undershooting can be seen frequently in the speed plots. With 80 messages being transmitted every second and the vehicle's response time at 300 ms, i.e., a total of 24 messages (or 23 redundant messages) are transmitted before the vehicle reacts. This is a bottleneck for the communication links as the vehicles cannot keep up with it. For both protocols, the PMs stabilize as per their position in the platoon. However, PM4 remains in an unstable state for a longer time owing to messages not being successfully transmitted during acceleration phase. Consequently, it starts later than PM3 and the increase in IVD can be seen (lower PDR at high distance). But, once the control messages are received by PM4 (peak of PM4 - IVD), it successfully re-joins the platoon and achieves stability.

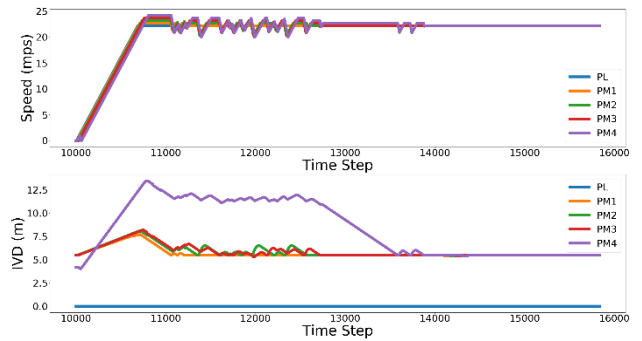

Fig 5: Platoon with IVD of 5m using DSRC protocol at a message rate of 80 Hz

Similar plots can be seen with a message rates of 40 Hz (see Figures 7 and 8) and 10 Hz (Figures 9 and 10). The only difference being that the redundant messages have dropped down to 11 (40 Hz) and 2 (10 Hz) before the vehicle reacts to the controller's commands. With lower number of redundant messages, the vehicles show lower instances of over- and undershooting.

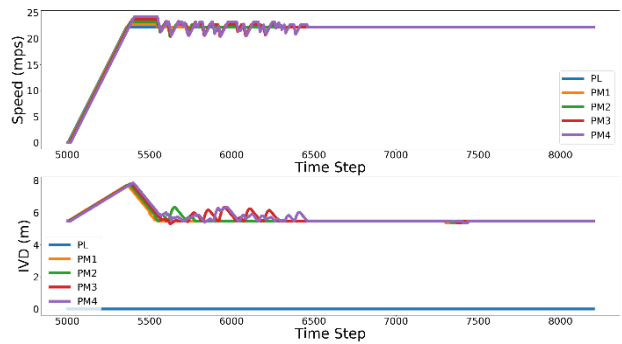

Fig 7: Platoon with IVD of 5m using DSRC protocol at a message rate of 40 Hz

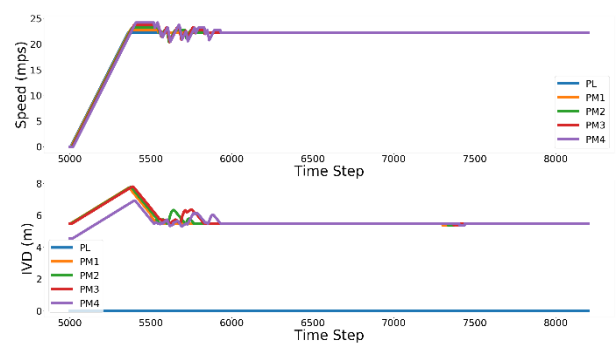

Fig 8: Platoon with IVD of 5m using LTE-V2V sidelink protocol at a message rate of 40 Hz

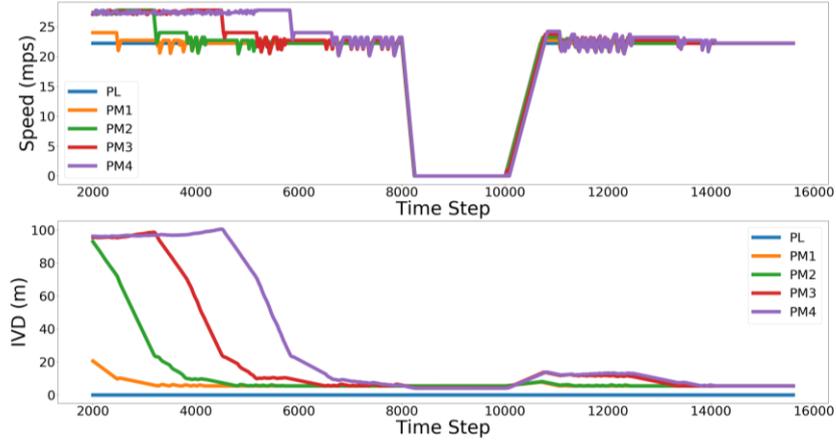

Fig 6: Complete simulation scenario showing platoon formation (2000 – 6500), emergency braking (8000) and acceleration (10000) phases for LTE-V2V sidelink for IVD of 5m at a message rate of 80 Hz

The time taken to achieve platoon stability in each of the scenarios can be found in Table 2. All times indicated in the table are averaged over several simulation runs, excluding the ones which involved collisions.

TABLE II. TIME TAKEN TO ACHIEVE PLATOON STABILITY

| Message Rate | DSRC | LTE-V2V sidelink |
|---|---|---|
| 80 Hz | ~39 seconds | ~31 seconds |
| 40 Hz | ~40 seconds | ~32 seconds |
| 10 Hz | ~27 seconds | ~25 seconds |

A notable difference in time to achieve stability can be found between message rates of 80/40 Hz and 10 Hz. This is primarily due to the fact that fewer messages are transmitted to stabilize the platoon. The reaction time of 300 ms proves to be a hindrance for higher message rates. Moreover, it can be seen that the LTE sidelink outperforms DSRC in every scenario. This can be accounted to the fact that the message delivery rates used for simulations are higher even at larger distances for the former than the latter. For example, at a distance of 75-100 m, the PDR value for DSRC is 91% [14] and for LTE sidelink is 97% [16].

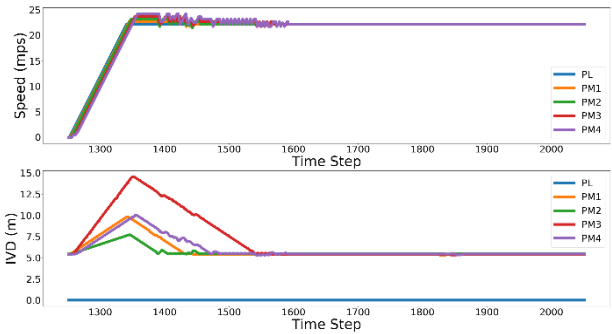

Fig 9: Platoon with IVD of 5m using DSRC protocol at a message rate of 10 Hz

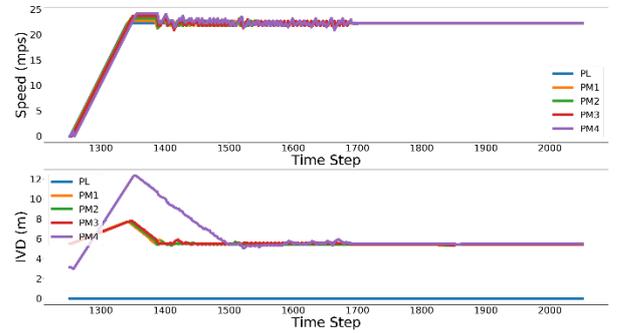

Fig 10: Platoon with IVD of 5m using LTE-V2V sidelink protocol at a message rate of 10 Hz

## VII. CONCLUSION

In this work, a controller was designed and implemented for a platooning use case, based on the principles of network-assisted CACC. Results indicate that the desired IVD and hence, the platoon stability is achieved over all message rates. Higher the desired IVD, the easier it is for the controller to achieve platoon stability. But, even the communication protocols play an important role, in terms of time taken, to achieve stability. LTE-V2V sidelink performs better than DSRC over all message rates. There have been simulation runs, where the controller has taken more time to stabilize the platoon, but platoon stability is achieved in every simulation. The bottleneck for this simulation is due to the limitation of vehicle reaction time. This means that even though a control update (message) is available, the vehicle is unable to react.

## VIII. FUTURE WORK

The controller used is a generic saw-tooth controller. Using a PID controller may provide smoother transitions for PMs and even a lower tolerance range can be implemented. This might lead to lower frequency of under- and over-shoots of vehicles during stabilizing. Also, changes in number of vehicles in the platoon or different vehicle reaction times can be tested to present a detailed model of the bottleneck. Also, the network protocols in this work are modelled on a high abstraction level. This method of implementation innately ensures that there is some semblance of guarantee for the message transmission to succeed. To overcome this problem, a more detailed network model, using a network simulator, including aspects such as latency, congestion, background traffic, resource allocation, etc. can be implemented.

IX. ACKNOWLEDGEMENT

A part of this work has been supported by the Federal Ministry of Education and Research of the Federal Republic of Germany (BMBF) in the framework of the project 16KIS0267 5GNetMobil. The authors alone are responsible for the content of the paper.